\documentclass[sigconf,screen]{acmart}
\AtBeginDocument{%
  }

\setcopyright{acmlicensed}
\copyrightyear{2026}
\acmYear{2026}
\setcopyright{cc}
\setcctype{by}
\acmConference[FORGE '26]{2026 IEEE/ACM Third International Conference on AI Foundation Models and Software Engineering}{April 12--13, 2026}{Rio de Janeiro, Brazil}
\acmBooktitle{2026 IEEE/ACM Third International Conference on AI Foundation Models and Software Engineering (FORGE '26), April 12--13, 2026, Rio de Janeiro, Brazil}
\acmPrice{}
\acmDOI{10.1145/3793655.3793740}
\acmISBN{979-8-4007-2477-0/2026/04}

\usepackage{color}
\usepackage{pifont}
\usepackage{listings}
\usepackage{subcaption}
\usepackage{multirow}
\usepackage{makecell}
\usepackage[linesnumbered,ruled,vlined]{algorithm2e}
\usepackage{enumitem}
\usepackage{url}
\usepackage{colortbl}

\newcommand{\ccSysName}[0]{SymTEE\xspace}
\newcommand{\ccBenchName}[0]{PartitioningE-Bench\xspace}
\newcommand{\ccTitle}[0]{Finding Missing Input Validation in TEEs via LLM-Assisted Symbolic Execution}

\newcommand{\blackding}[1]{\ding{\numexpr181+#1\relax}}

\newcommand{\tabincell}[2]{
	\begin{tabular}{@{}#1@{}}
		#2
	\end{tabular}
}

\begin{document}

\title{\ccTitle}


\author{Chengyan Ma}
\affiliation{%
  \institution{Singapore Management University}
  \country{Singapore}
}
\author{Jieke Shi}
\affiliation{%
  \institution{Singapore Management University}
  \country{Singapore}
}
\author{Ruidong Han}
\affiliation{%
  \institution{Singapore Management University}
  \country{Singapore}
}
\author{Ye Liu}
\affiliation{%
  \institution{Singapore Management University}
  \country{Singapore}
}
\author{Yuqing Niu}
\affiliation{%
  \institution{Singapore Management University}
  \country{Singapore}
}
\author{David Lo}
\affiliation{%
  \institution{Singapore Management University}
  \country{Singapore}
}

\renewcommand{\shortauthors}{Ma et al.}

\begin{abstract}
Trusted Execution Environments (TEEs) provide hardware-enforced isolation that protects sensitive code and data from untrusted software. Despite their strong security guarantees, analyzing TEE applications remains challenging due to the high cost and complexity of configuring complete TEE build and runtime environments, as well as the limited observability imposed by hardware isolation.
This paper presents \ccSysName, a novel large language model (LLM)-assisted symbolic execution framework for detecting missing input validation issues in TEE applications without requiring real TEE setups. \ccSysName begins by leveraging Abstract Syntax Tree (AST) analysis to extract TEE code slices that may lack sufficient input validation, and then employs an LLM (GPT-5 in our case) to automatically convert the extracted slices into KLEE-compatible harness programs containing lightweight mock execution environments for symbolic analysis.
Evaluations on 26 vulnerabilities (11 real-world and 15 synthetic) show that \ccSysName achieves 100\% precision and 92.3\% recall in detecting missing input validation vulnerabilities while incurring an average analysis cost of only \$0.05. These results demonstrate the effectiveness and practicality of \ccSysName's pioneering paradigm of LLM-assisted symbolic execution, where LLMs autonomously generate mock environments to enable automated security analysis without complex setup, providing a more accessible and scalable framework for trusted computing systems.
\end{abstract}

\maketitle

\section{Introduction}
\label{sec:intro}
\begin{figure}[t]
    \centering
    \includegraphics[width=0.95\linewidth]{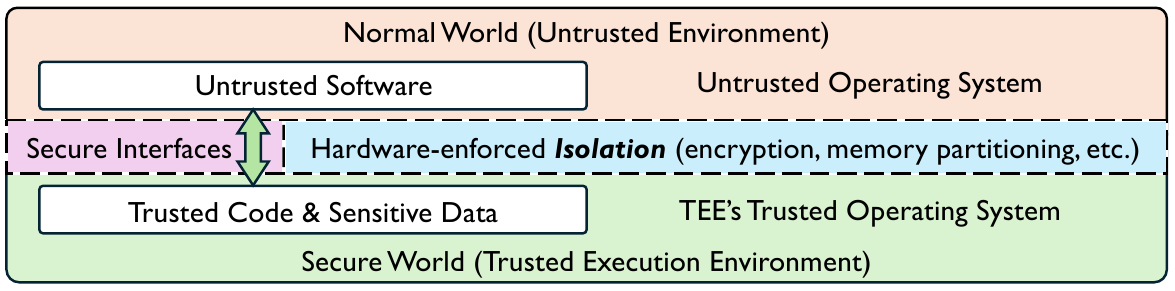}
    \caption{Architecture of a Trusted Execution Environment (TEE), which is isolated from the normal world and can only be accessed via secure APIs.}
    \label{fig:tee}
\end{figure}

A Trusted Execution Environment (TEE) is a segregated area within a processor and its memory that isolates sensitive code and data from potentially compromised software or operating systems~\cite{9041685,li2023survey,7345265}, thereby ensuring the confidentiality and integrity of critical computations and assets like cryptographic keys~\cite{6296112} and biometric information~\cite{Anasuri2023,9041685}. Major hardware vendors, including Intel and ARM, have developed TEE infrastructures, with Intel Software Guard Extensions (SGX)~\cite{intelsgx} and ARM TrustZone~\cite{armtrustzone} being two widely-adopted implementations. As shown in~\autoref{fig:tee}, they typically divide the computing environment into a normal world and a secure world (TEE). The normal world hosts untrusted components that cannot access or tamper with data in the secure world, which is protected by hardware-enforced isolation mechanisms like memory partitioning and encryption, and can only be accessed via secure application programming interfaces (APIs) provided by hardware vendors. TEEs have become a cornerstone of trustworthy computing, supporting applications such as mobile payments~\cite{9514868,10.1145/3313808.3313810,10.1007/978-3-319-45871-7_9}, digital identity management~\cite{9367444,li2023survey}, and many other security-critical services that collectively safeguard assets and user data.

\begin{figure}[t!]
\centering
\begin{lstlisting}
// GitHub Project: shuaifengyun/basicAlg_use (commit 327f23d)
// File: core/tee/crypto_ta_pbkdf2.c
@@ -261,6 +261,9 @@ void g_CryptoTaPbkdf_PBKDF2(..., int dkLen)
     // Improper TEE usage: missing input validation of data from untrusted normal world
-    TEE_MemMove(output, resultBuf, dkLen);
     // Fix: Added input validation to prevent buffer overflow
+    if (dkLen > 512)
+        return TEE_ERROR_BAD_PARAMETERS;
+    TEE_MemMove(output, resultBuf, dkLen);
     return TEE_SUCCESS;
\end{lstlisting}
\caption{Example of missing input length validation in a memory copy operation from the untrusted normal world to a fixed-size buffer within the TEE, found in a cryptographic library on GitHub (\texttt{shuaifengyun/basicAlg\_use}). Our \ccSysName successfully detected this issue and suggested a fix.}
\label{fig:example}
\end{figure}

However, developing secure TEE-based applications is challenging. Similar to the insecure practices often observed in cryptographic and other security-critical contexts~\cite{10.1145/2884781.2884790,10.1145/3735968,10.1145/3674805.3686685}, many developers with limited security expertise may misuse or insecurely apply the secure APIs provided by hardware vendors when building TEE applications, thereby undermining security guarantees or even introducing new attack surfaces. A common class of vulnerabilities in TEE applications arises from insufficient validation and sanitization of input data received from the potentially compromised normal world, including improper or missing checks on data size, format, or validity~\cite{ma2025ditingstaticanalyzeridentifying}. Such flaws can lead to severe memory corruption issues such as buffer, stack, and heap overflows~\cite{10.1145/3373376.3378486,ma2025ditingstaticanalyzeridentifying}.
\autoref{fig:example} illustrates this issue with a vulnerability found in a TEE-based cryptographic library on GitHub, where the developer failed to validate the input length of parameters received from the normal world. This omission can result in buffer overflows, turning the TEE from a security boundary into an exploitable target, where attackers can craft malicious inputs that corrupt memory and potentially compromise the secure environment.

Detecting missing input validation issues is challenging because such bugs often occur only under specific input conditions and execution paths (e.g., when input sizes exceed buffer limits), making them difficult for static analysis to reliably confirm~\cite{10.1145/3373376.3378486,ma2025ditingstaticanalyzeridentifying} and often requiring dynamic analysis to capture their runtime semantics. However, the unique architecture of TEEs introduces multiple obstacles for dynamic analysis techniques such as fuzzing and symbolic execution, which have proven effective for conventional software~\cite{10.1145/3623375,10.1145/3182657}. Concretely, TEEs are accessible only through vendor-defined secure interfaces that common security-oriented fuzzers~\cite{10858174} or symbolic executors~\cite{10.5555/1855741.1855756} do not natively support~\cite{10.1145/3611668,9041685,ma2025ditingstaticanalyzeridentifying}, forcing researchers to build custom harnesses and loaders that are often tedious and error-prone. Executing real TEEs further requires a specialized toolchain and environment for cross-compilation and signing on hardware with TEE support~\cite{li2023survey}, so even running a simple ``hello-world'' test case demands substantial setup effort and may be infeasible when appropriate hardware is unavailable. Moreover, TEE isolation prevents the normal world from observing or instrumenting the secure world, rendering standard debugging methods ineffective because crashes or coverage data cannot be collected without modifying the TEE or its APIs. Collectively, these constraints cause most off-the-shelf dynamic analysis tools to fail when analyzing TEE applications, thereby overlooking subtle input-validation flaws hidden within the secure world.

In this paper, we present \ccSysName, a Large Language Model (LLM)-assisted symbolic execution framework for detecting missing input validation issues in TEE applications. The core idea of \ccSysName is to tackle the aforementioned challenges by leveraging LLMs to automatically generate mock environments that emulate TEE runtimes with minimal stub dependencies, eliminating the need for full TEE setups or specialized hardware.

\ccSysName starts with performing Abstract Syntax Tree (AST)-based analysis to extract code slices that may lack input validation, enabling focused analysis on high-risk code regions. Then, it uses an LLM (OpenAI GPT-5 in our case) to expand each code slice into a complete program with minimal stubs compatible with KLEE~\cite{10.5555/1855741.1855756}, a dynamic symbolic execution engine, allowing symbolic execution to run locally without instrumenting a real TEE. Finally, \ccSysName applies KLEE to explore execution paths and identify inputs that violate the assertion oracles generated in the previous step by the LLM, which encode input validation checks.

We evaluate \ccSysName on a benchmark of 26 vulnerabilities, including 11 real-world cases from GitHub projects and 15 synthetic ones from the recent benchmark \ccBenchName~\cite{ma2025ditingstaticanalyzeridentifying}. \ccSysName achieves a precision of 100\% and a recall of 92.3\% in detecting missing input validation issues, with each analysis requiring, on average, 5,931 tokens (costing approximately 0.05 USD). These results demonstrate that \ccSysName is both effective and practical for detecting TEE vulnerabilities and represent an early yet promising step toward LLM-assisted symbolic execution based on automatically generated mock environments, paving the way for more accessible and efficient security analysis in trusted computing systems and other constrained environments.

\noindent\textbf{Our contributions:}
\begin{itemize}[leftmargin=*,nosep]
    \setlength{\topsep}{0pt}
    \setlength{\parsep}{0pt}
    \item We propose the first LLM-assisted symbolic execution approach for detecting missing input validation issues in TEE applications, without requiring complex runtime setups or hardware support.
    \item We perform a comprehensive evaluation on 26 vulnerabilities, including both real-world and synthetic programs, demonstrating that \ccSysName achieves 100\% precision and 92.3\% recall.
    \item As an emerging line of work, we provide insights for future research and outline potential extensions, like enhancing \ccSysName with specialized LLMs and broadening its vulnerability coverage.
\end{itemize}

\section{Preliminaries and Related Work}

\subsection{Preliminaries}
\noindent\textbf{Trusted Execution Environments (TEEs).}  As introduced in \autoref{sec:intro}, TEEs are isolated execution regions within a processor that protect sensitive code and data from untrusted software. This isolation is enforced by hardware mechanisms, making program analysis and testing of TEE applications particularly challenging because internal states and runtime behaviors are largely inaccessible to external tools. Moreover, setting up a TEE environment also requires complex configurations such as cross-compilation, secure boot, and binary signing with vendor-specific toolchains~\cite{li2023survey,10.1145/3456631}. These factors collectively hinder the use of traditional dynamic analysis techniques such as fuzzing and symbolic execution for vulnerability detection in TEE applications, highlighting the need for alternative approaches like \ccSysName that can perform analysis locally without relying on actual TEE hardware or runtime environments.

\begin{figure*}[t!]
    \centering
    \includegraphics[width=0.89\linewidth]{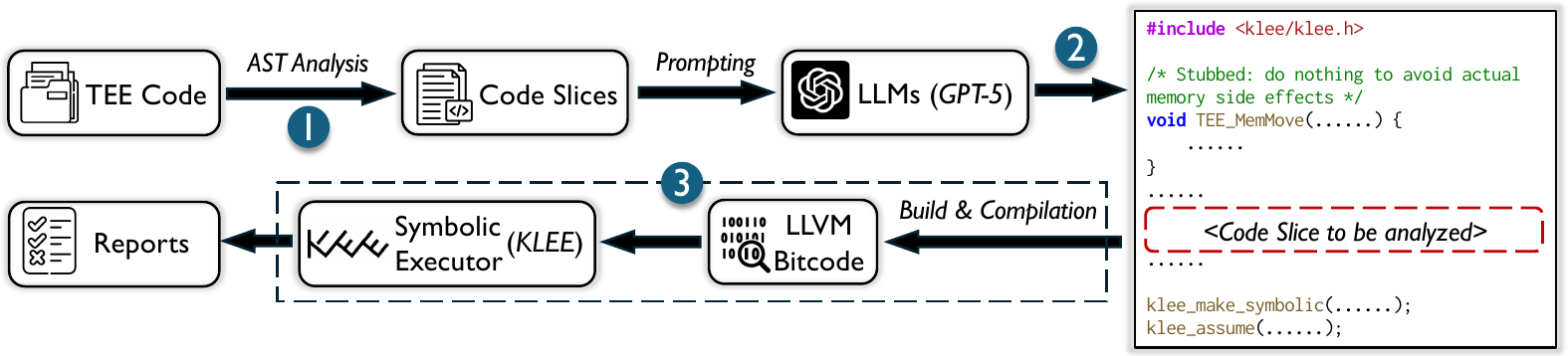}
    \caption{Overview of the \ccSysName\ workflow consisting of three stages: \blackding{1} AST analysis for extracting code slices; \blackding{2} LLM-assisted generation of mock environments; and \blackding{3} dynamic symbolic execution for vulnerability detection.}
    \label{fig:workflow}
\end{figure*}

\vspace{0.1cm}
\noindent\textbf{Symbolic Execution.} Symbolic execution~\cite{10.1145/3182657,10.5555/1855741.1855756,10.1145/2110356.2110358} is a program analysis technique that treats inputs as symbolic variables instead of concrete values, enabling simultaneous exploration of multiple execution paths. For each path, it generates logical constraints that describe the conditions under which that path is taken. By solving these constraints with Satisfiability Modulo Theories (SMT) solvers~\cite{10.1145/1995376.1995394}, concrete inputs can be derived that trigger specific program behaviors, facilitating targeted testing and vulnerability discovery.
Tools such as KLEE~\cite{10.5555/1855741.1855756}, S2E~\cite{10.1145/2110356.2110358}, and angr~\cite{7546500} have been widely adopted in security analysis to identify bugs and verify correctness. In our work, since TEE applications are primarily written in C/C++, we use KLEE, one of the most widely used symbolic execution engines for C/C++ programs, to systematically explore code paths and detect missing input validation vulnerabilities.

\subsection{Related Work}
Recent research has explored methods for identifying vulnerabilities in TEE applications.
DITING~\cite{ma2025ditingstaticanalyzeridentifying} is a static analyzer tailored for TEE projects that tracks data flow between the normal world and the TEE to detect insecure patterns such as unencrypted data exchange. Other efforts~\cite{10.1145/3456631,9152801,10.1145/3407023.3407072} have empirically studied and characterized common TEE weaknesses, offering valuable guidance for rule-based static analysis.
Beyond static inspection, dynamic analysis has also been investigated. COIN~\cite{10.1145/3373376.3378486} introduces a concolic execution framework for testing Intel SGX, while PARTEMU~\cite{247658} employs QEMU~\cite{10.5555/1247360.1247401} to emulate TrustZone for dynamic testing. Fuzzing-based approaches such as TEEzz~\cite{10179302}, EnclaveFuzz~\cite{DBLP:conf/ndss/ChenLMLC024}, and TEEFuzzer~\cite{DBLP:journals/fgcs/DuanFZDSCC23} aim to generate valid inputs and improve code coverage through techniques like input format inference and coverage-guided mutation.
However, all existing dynamic methods require dedicated TEE compilation and runtime environments, inevitably facing challenges of complex setup and reliance on TEE-enabled hardware.

\section{Methodology}
As illustrated in~\autoref{fig:workflow}, \ccSysName contains 3 steps:
\blackding{1} Analyzing the TEE code via its AST to extract all the code slices that may have vulnerabilities.
\blackding{2} We use LLM to expand the code slice into complete code that can be analyzed by KLEE (a dynamic symbolic execution engine).
\blackding{3} After compiling the code to LLVM bitcode, we can use dynamic symbolic execution to analyze whether a vulnerability actually exists.
We elaborate on these steps as follows.

The first step performs AST analysis to identify code slices that may lack input validation.
Since TEE memory copy operations must use fixed APIs provided by vendors (e.g., \texttt{TEE\_Move(dest, src, len)}), \ccSysName detects such operations and tracks data flow through the AST to locate the field that specifies the size of the destination buffer \texttt{dest}.
It then checks whether this size field is compared against the copy length (\texttt{len}) to ensure proper boundary validation.
If no such comparison is found, the operation is flagged as potentially vulnerable to buffer overflow.
At this stage, \ccSysName extracts the entire function containing the suspicious operation as a code slice for subsequent dynamic analysis.

Because the extracted TEE code slices depend on vendor libraries and hardware, they cannot be compiled or executed on a normal host directly. In Step \blackding{2}, we therefore leverage an LLM to generate a mock environment for each slice. The mock environment has two goals. First, it supplies minimal stub implementations for external dependencies (e.g., \texttt{TEE\_*} APIs and types such as \texttt{TEE\_Param}) so the slice compiles as a standalone C program. Second, it constructs a KLEE-compatible security harness (\texttt{int main(void)}) for dynamic analysis: the harness marks attacker-controlled inputs (for example, buffer length \texttt{len} or data content) symbolic via \texttt{klee\_make\_symbolic} and constrains them with \texttt{klee\_assume}. It also embeds a security oracle (using \texttt{klee\_assert}) that checks function return values or an instrumentation flag (e.g., \texttt{g\_checked}) to determine whether the code correctly rejects malicious inputs before performing memory operations. The result is a single, self-contained C file suitable for compilation to LLVM bitcode and subsequent symbolic execution in Step \blackding{3}.
\autoref{fig:klee_code} illustrates a KLEE-compatible program synthesized from a vulnerable TEE slice: the LLM embeds the original \texttt{produce} function (with the unsafe \texttt{TEE\_MemMove}), generates minimal type and API stubs so the slice compiles, and synthesizes a \texttt{main} harness that makes \texttt{attacker\_buf} and \texttt{attacker\_size} symbolic (constrained with \texttt{klee\_assume}). A \texttt{klee\_assert} oracle flags when \texttt{g\_checked} is unset and the condition \texttt{attacker\_size > 512UL} is met, allowing KLEE to find concrete inputs that violate the check.

\begin{figure}[t!]
\centering
\begin{lstlisting}
#include <klee/klee.h>

typedef struct { void* buffer; unsigned long size; } memref_t;
typedef struct { memref_t memref; } TEE_Param;
volatile int g_checked = 0;

void TEE_MemMove(void* dest, const void* src, size_t n) {
    (void)dest; (void)src; (void)n;
    /* Stubbed: do nothing to avoid actual memory side effects */
}

void produce(TEE_Param params[4]) {
    char str[512];
    TEE_MemMove(str, params[0].memref.buffer, params[0].memref.size);
}

int main(void) {
    TEE_Param params[4];

    char buf[4096];
    unsigned long size;
    klee_make_symbolic(&size, sizeof(size), "size");
    klee_assume(size <= 4096UL);

    params[0].memref.buffer = buf;
    params[0].memref.size = size;

    produce(params);

    if (size > 512UL) {
        klee_assert(g_checked && "Missing input validation");
    }
    ......
}
\end{lstlisting}
\caption{An example of LLM-generated KLEE harness.}
\label{fig:klee_code}
\end{figure}

Once the LLM-generated KLEE harness is produced, we compile it to LLVM bitcode and pass it to KLEE in Step \blackding{3}. KLEE explores execution paths driven by the symbolic inputs (e.g., \texttt{attacker\_size}, \texttt{attacker\_buf} in the example) and searches for concrete values that violate the \texttt{klee\_assert} oracle. If KLEE successfully finds a path that triggers the assertion (i.e., a path where inputs of invalid size, range, or format are processed without proper validation, leading to memory corruption), the slice is reported as a true vulnerability. \ccSysName then generates a report describing the execution path, the concrete inputs, and the location of the missing validation.
\begin{table}[t!]
    \centering
    \caption{Effectiveness evaluation and GPT-5 token usage of \ccSysName. Vul is the number of issues, N is the number of detection results, and TP is the number of true positives. P and R indicate the precision and recall, respectively.}
    \label{tab:eva}
    \footnotesize
    \setlength{\tabcolsep}{0.7mm}
    \begin{tabular}{ccccc}
    \toprule
         \textbf{Project} & \textbf{\#Vul} & \textbf{\tabincell{c}{Detection Results\\(N/TP)}} & \textbf{\tabincell{c}{Token Usage \\ (On Avg.)}} & \textbf{\tabincell{c}{Token Cost \\ (On Avg.)}} \\
         \midrule
         \multicolumn{5}{c}{Result on \ccBenchName~\cite{ma2025ditingstaticanalyzeridentifying}}\\
         \hline
         \ccBenchName & 15 & 13/13 & 4,223 & \$0.03 \\
         \midrule
         \multicolumn{5}{c}{Result on real-word projects}\\
         \hline
         optee-sdp & 9 & 9/9 & 7,215 & \$0.06 \\
         basicAlg\_use & 2 & 2/2 & 6,354 & \$0.05 \\
         \midrule
         Total & \multirow{2}{*}{26} & 24/24 & \multirow{2}{*}{5,931} & \multirow{2}{*}{\$0.05} \\
         P(\%)/R(\%)* &  & 100/92.3 & & \\
    \bottomrule
    \multicolumn{5}{l}{* P(\%) = \#TP / \#N, R(\%) = \#TP / \#Vul.}
    \end{tabular}
\end{table}

\section{Evaluation}
\label{sec:eva}

We evaluate \ccSysName on a benchmark comprising 26 memory operations without input validation. This benchmark is constructed from two distinct sources:
\begin{itemize}[leftmargin=*,nosep]
    \setlength{\topsep}{0pt}
    \setlength{\parsep}{0pt}
    \item \textbf{Synthetic Test Cases}: 15 unsafe memory operation vulnerabilities chosen from the \ccBenchName test suite~\cite{ma2025ditingstaticanalyzeridentifying}, a collection of security flaws originally created by manual injection into the function bodies with diverse coding styles.
    \item \textbf{Real-World Projects}: 11 real-world vulnerable code fragments collected from open-source TEE projects on GitHub (\textit{optee-sdp} and \textit{basicAlg\_use}).
\end{itemize}
By incorporating both complex vulnerabilities sourced from real-world projects and synthetic flaws from a controlled test suite, the evaluation of \ccSysName ensures broad coverage and high representativeness. 
\ccSysName is implemented in Python with GPT-5 and evaluated on a server running Ubuntu 24.04, equipped with a 48-core 2.3 GHz AMD EPYC 7643 processor and 512 GB RAM.

\autoref{tab:eva} presents the results. Out of the 26 known vulnerabilities in the benchmark, \ccSysName successfully detected 24 issues, achieving a recall of 92.3\% and a precision of 100\%.
The two false negatives were primarily due to the limitations of the initial static analysis phase (\blackding{1}), which failed to extract the relevant code slices containing the vulnerable memory operation.
For all 24 code slices that were successfully extracted by the static analysis, the subsequent dynamic symbolic execution phase (\blackding{3}) correctly identified the missing input validation vulnerability.
We also measured the token usage and cost of GPT-5. On average, analyzing each vulnerability required about 5,931 tokens, resulting in an average token cost of \$0.05 per vulnerability.
Overall, these results demonstrate that \ccSysName can reliably generate compilable mock environments and KLEE harnesses, accurately identify unsafe memory operations through symbolic execution, and achieve strong cost-effectiveness in TEE vulnerability detection.

\section{Future Plans}

\textbf{Strengthening AST Analysis and LLM Specialization.}
Our failure-case analysis (\autoref{sec:eva}) suggests that \ccSysName could benefit from a more precise AST-based slicer to better capture potentially vulnerable code slices. Future work can improve the slicing heuristics and integrate interprocedural data-flow analysis to enhance the slicer’s accuracy.
Moreover, the effectiveness of \ccSysName also depends on the LLM’s ability to generate correct mock environments and KLEE harnesses. Although general-purpose models performed well on our benchmark, they may struggle with real-world TEE applications that involve complex APIs or uncommon features. Future research can investigate training or fine-tuning LLMs on TEE codebases to strengthen their understanding of TEE semantics and enable the generation of more reliable mock environments.

\vspace{0.1cm}
\noindent\textbf{Benchmark and Scope Extensions.}
Currently, \ccSysName specializes in detecting missing input validation vulnerabilities in TEE applications. However, TEEs also suffer from other security issues, such as improper cryptographic usage~\cite{ma2025ditingstaticanalyzeridentifying,9152801}. We plan to extend the current benchmark to include more diverse vulnerabilities and real-world TEE projects, allowing a broader and more realistic evaluation of \ccSysName's versatility.
Moreover, our proposed paradigm of LLM-assisted symbolic execution can be generalized to other domains, such as embedded systems~\cite{10.1145/3613905.3650764,10.1145/3538644}, which also involve complex toolchains and hardware-reliant debugging processes. We aim to adapt \ccSysName to these environments to make security analysis more accessible across a wider range of critical systems.

\vspace{0.1cm}
\noindent\textbf{Trust and Synergy with Developers.}
Automated vulnerability detection assisted by LLMs, including our approach, currently involves limited interaction with developers, which poses challenges for building trust and achieving effective collaboration. This issue has been increasingly discussed in the context of AI-assisted software engineering~\cite{10.1145/3712003,10449668,10.1145/3708525}. Future research should explore mechanisms to promote closer collaboration, for example, by clarifying the impact of mock environments and enabling developers to review or refine them to ensure safer execution. Strengthening this trust and interaction will help LLM-powered symbolic executors evolve into reliable collaborators, aligning with the vision of trustworthy and synergistic AI in software engineering~\cite{10449668}.

\section{Conclusion}
This paper presents \ccSysName, a novel LLM-assisted symbolic execution approach for detecting missing input validation issues in TEE applications.
\ccSysName leverages LLMs to automatically generate mock environments and KLEE-compatible security harnesses, eliminating the need for complex TEE runtime setups or specialized hardware.
Experiments on 11 real-world vulnerabilities and 15 synthetic cases show that \ccSysName achieves 100\% precision and 92.3\% recall at an average cost of about \$0.05 per analysis, demonstrating both effectiveness and practicality in uncovering input validation flaws.
Our study also points to opportunities for extending LLM-assisted symbolic execution to broader classes of TEE vulnerabilities and other constrained environments like embedded systems.
As an emerging result with substantial potential impact, \ccSysName embodies an early yet promising paradigm where LLMs autonomously generate mock environments to enable automated security analysis without complex setup, paving the way for more accessible and scalable analysis for trusted computing systems.

\begin{acks}
This research/project is supported by the National Research Foundation, Singapore, and the Cyber Security Agency of Singapore under its National Cybersecurity R\&D Programme (Proposal ID: NCR25-DeSCEmT-SMU). Any opinions, findings and conclusions or recommendations expressed in this material are those of the author(s) and do not reflect the views of the National Research Foundation, Singapore, and the Cyber Security Agency of Singapore.
\end{acks}

\section*{Data Availability}
A replication package, including the implementation
of \ccSysName and detailed instructions for running it, is available
at: \url{https://github.com/CharlieMCY/SymTEE}.

\bibliographystyle{ACM-Reference-Format}
\bibliography{sample-base,software}


\end{document}